\documentstyle[aps,epsf,epsfig]{revtex}
\begin{document}
\input epsf
\title{Classical  Electrodynamics and the Quantum Nature of Light}
\author{Manoelito M de Souza}
\address{Universidade Federal do Esp\'{\i}rito Santo - Departamento de
F\'{\i}sica\\29065.900 -Vit\'oria-ES-Brasil}
\date{\today}
\maketitle
\begin{abstract}
\noindent A review of old inconsistencies of   Classical
Electrodynamics (CED) and of some new ideas that solve them is presented. Problems with causality violating solutions of the wave equation and of 
the electron equation of motion, and problems with the non-integrable singularity of its self-field  energy tensor are well known. The correct interpretation of the two (advanced and retarded) Lienard-Wiechert 
solutions are in terms of creation and annihilation of particles in 
classical physics. They are both retarded solutions. Previous work on the short distance limit of CED of a spinless point electron are based on a faulty assumption which causes the well known inconsistencies of the 
theory: a diverging self-energy (the non-integrable singularity of its self-field  energy tensor) and a causality-violating third order equation 
of motion (the Lorentz-Dirac equation). The correct assumption fixes these problems without any change in the Maxwell's equations and let exposed, in the zero-distance limit, the discrete nature of light: the flux of energy from a point charge is discrete in time. CED cannot have a true equation of motion but only an effective one, as a consequence  of the intrinsic 
meaning of the Faraday-Maxwell concept of field that does not correspond to the classical description of photon  exchange, but only to the smearing of its effects in the space around the charge. This, in varied degrees, is transferred to QED and to other field theories that are based on the same concept of fields as space-smeared interactions.
\end{abstract}
\begin{center}
PACS numbers: $03.50.De\;\; \;\; 11.30.Cp$
\end{center}

\begin{section}
{INTRODUCTION}
\end{section}

\noindent Classical and quantum physics are considered to involve sharply distinct concepts and kinds of  theories. Classical physics represents an approximation of the more refined and closer to a true description of the 
world which is supposedly done by quantum physics.
The short-distance limit of both has always been plagued by unsurmountable problems, which in CED are attributed to the assumed pointlike nature of 
the electron. Assuming a finite non zero dimension for the electron  
brings, however, more problems than solutions.  The blame is not on the pointlike electron but on an incorrect approach on taking the theory zero distance limit. A more careful approach will free the theory of theses problems and will reveal some quantum aspects, up to now unsuspected, in a classical theory.\\
\noindent CED of a point electron is based on the
Lienard-Wiechert solution (LWS);  its many old and unsolved problems
\cite{Rohrlich,Jackson,Parrot} make of it a non-consistent theory.  The Lienard-Wiechert advanced solution represents itself a causality problem 
that has required some, at least verbal, efforts to be circumvented. One 
must also mention  the field singularity or the self-energy problem;  the non-integrable singularities of its energy tensor;  the causality-violating behaviour of solutions of the Lorentz-Dirac equation \cite{Eliezer,Teitelboim,Rowe,Marx,Lozada};
etc. It will be shown here that the solution to these problems is
connected to a more strict implementation of causality (extended 
causality), already present, although not yet recognized, in the Lienard-Wiechert solution. In section II the notation is defined in a 
brief review of the standard interpretation of the two (advanced and retarded) LWS.  Causality can be seen as a restriction to access to 
regions of the spacetime manifold, as discussed in section III, where
 the notion of extended causality is introduced; it allows a new interpretation of the two LWS in terms of creation and annihilation of classical particles. The notion of a classical photon is introduced. In section IV the singularities and non-integrability of the electron 
self-field energy tensor as they are described in the literature are 
reviewed and discussed.  It is then pointed that they are all 
consequences of using an implicit assumption about the zero distance limit that will be proved faulty in the following section.
Section V shows how to correctly take the zero distance limit in CED and to give to it a consistent physical interpretation. The anticipated 
recognition in the classical theory of the actual quantum nature of the electromagnetic radiation is necessary for having a clear physical picture  behind these new mathematical results. Some algorithms, that will be used 
in the rest of the paper for taking the zero distance limit, are presented in section VI.  In section VII, while searching for an electron ``equation of motion", it is confirmed by an explicit direct calculation that the old problem of singularity and non-integrability of the electron self-field energy tensor has vanished just with correctly taking the zero-distance limit. No 
tampering with the maxwell's equations and with the energy tensor is ever done in this paper. All that is allowed is a possible reinterpretation of their physical meaning. The most remarkable new feature is that the energy flux from a point charge is discrete in time, which  requires an interpretation of light in terms of discrete emission of pointlike objects (classical photons) and a revision of the physical meaning of the Gauss's 
law and of the Faraday-Maxwell concept of field. This  will be done in the last section. The electron ``equation of motion", derived in section VIII, does not have the problematic Schott term but it is just an effective equation. This is a consequence of the bi-local character of the LWS, as it depends on two possibly far apart points: the point where the signal is defined, and the point, in the source worldline, where it was created. It 
is argued then that CED cannot produce a true equation of motion for its 
sources as far as its formulation is based on the Faraday-Maxwell concept of fields. Section IX is included like an appendix of section VIII for showing an alternative calculation that enlightens its physical meaning. The paper 
is concluded in section X with a summary and a discussion of the physical content of the Maxwell-Faraday concept of field upon which the modern field theory is entirely based. Gauss's law is not compatible with the vision of 
a classical field in terms of exchange of discrete objects (classical photons) unless the fields represent rather space average effects taken 
over a period of time larger than the time interval among the photon emissions. The Faraday-Maxwell concept of field, which is based on the validity of the Gauss's law, represents the smearing over the space surrounding the charge of the effects of the exchanged photons. The field singularity at the charge position is a reflection of this smearing process or the field space-average character.

\begin{section}
{THE LIENARD-WIECHERT SOLUTIONS}
\end{section}

\noindent The retarded Lienard-Wiechert potential
\begin{equation}
\label{LWS}
 A(x)=\frac{V}{\rho}{\Big|}_{{\tau}_{ret}},\;\;$for$\;\;\rho>0,
\end{equation}
is a (the retarded one) solution to the wave equation
\begin{equation}
\label{BoxA}
\Box A(x)=4\pi J(x)
\end{equation}
 and to the gauge condition,
\begin{equation}
\label{dA}
\partial.A\equiv\frac{\partial A^{\mu}}{\partial x^{\mu}}=0,
\end{equation}
 where J, given by
\begin{equation}
\label{J}
J(x)=\int d\tau V\delta^{4}[x-z(\tau)],
\end{equation}
is the  current  for  a  point electron that describes a given trajectory
$z(\tau)$, parameterized by its proper-time $\tau;$ $V=\frac{dz}{d\tau}.$ 
The
electron charge and the speed of light are taken as 1.
\begin{equation}
\label{rho}
\rho:= -V_{\alpha}R^{\alpha}=-V.\eta.R=-V.R,
\end{equation}
 where $\eta$ is the  Minkowski metric tensor diag(-1,1,1,1),  and $R:=
x-z(\tau)$. $\rho$ is the invariant distance (in the charge rest frame) between
$z(\tau_{ret}),$ the position of the charge at the retarded time, and x, its
self-field event (See figure 1).
The constraints
 \begin{equation}
\label{lcone}
R^{2}=0,
\end{equation}
and
\begin{equation}
\label{rett}
R^{0}>0,
\end{equation}
 must be satisfied.
The constraint
$R^{2}=0$ requires that x and $z(\tau)$ belong to a same light-cone; it has two
solutions, $\tau_{ret}$ and $\tau_{adv}$, which are, respectively, the points where J
intercepts the past and the future light-cone of x (see figure 1). The
retarded solution describes a signal emitted at $z(\tau_{ret})$ and that is
being observed at x, with $x^{0}>z^{0}(\tau_{ret})$, while the advanced
solution also observed at x, {\it will be} emitted in the future, at
$z(\tau_{adv})$, with $x^{0}>z^{0}(\tau_{adv})$. $R^{0}>0$ is a restriction to
the retarded solution (\ref{LWS}) as it excludes the causality violating
advanced solution, and justifies the restriction ${\big|}_{{\tau}_{ret}}$ in
(\ref{LWS}). But this is not the only available interpretation; it will be shown below another one that does not have problems with causality violation and that, remarkably, allows the description of particle  creation and annihilation still in a classical physics context.\\

\begin{section}
{ Causality and spacetime geometry}
\end{section}

There is a well known geometric and physical interpretation of the  constraint
(\ref{lcone}). $R^{2}=0$ assures that $A(x)$ is a signal that
propagates with the speed of light, on a light-cone; in field theory it
corresponds to the implementation of the so called {\it local causality}: only
points inside or on a same light-cone can be causally connected. It defines for
a physical object, at a point, {\it its physical spacetime}, that is the
regions of the space-time manifold that it can have access to. In the literature only (\ref{lcone}) is clearly associated to the notion of causality but this is not enough because A(x), in CED, is just an ancillary intermediary step to the Maxwell stress tensor, to whose components, the electric and magnetic fields, are attributed the physical meaning of force carriers. So, it is necessary to consider variations of (\ref{LWS}) and, therefore of (\ref{lcone}). \\
The constraint (\ref{lcone})
 must be considered in the neighbourhoods of x and of z: $x+dx$ and
$z(\tau_{ret}+d\tau)$ must also belong to a same light-cone.  A differentiation
 of (\ref{lcone}) ($R.dR=0\rightarrow R.(dx-Vd\tau)=0\rightarrow R.dx+\rho
d\tau=0$) generates the  constraint
\begin{equation}
\label{RdR}
d\tau+K.dx=0,
\end{equation}
 where K, defined for $\rho>0$, by
\begin{equation}
\label{K}
K:=\frac{R}{\rho},
\end{equation}
is a null 4-vector, $K^{2}=0$, and represents a light-cone generator, a tangent
to the light-cone.
The constraint (\ref{RdR}) is a condition of consistency of (\ref{lcone}). It defines a family of hyperplanes tangent to the light-cone defined by $R^{2}=0$.   Together, these two
constraints require that x and $z(\tau_{ret})$ belong to a same straight line, generator of the lightcone with vertex at the point x. This generator is the one tangent to $K^{\mu},$ according to (\ref{K}), and  orthogonal to
$K_{\mu},$ according to (\ref{RdR}): $K^{\mu}K_{\mu}=0.$ At the vertex of a lightcone the generators come in pairs: $K^{\mu}:=(K^{0},{\vec K})$  and ${\bar
K}^{\mu}:=(K^{0},-{\vec K}).$ \\
Together, (\ref{lcone})  and (\ref{RdR}) produce a much more restrictive
causality constraint: a free massless physical object is restricted to remain on its  light-cone generator (labelled by K). This is a very powerful restriction and changes radically the nature of field theory. One is not dealing anymore with distributed fields defined on the whole lightcone but with a localized object (its $(t=const)$-intersection is not a 2-sphere but a point!) defined on a light-cone generator. Or, in other words, the part of a wavefront of $A(x)$ that moves along a light-cone generator must remain on this same generator. This is in direct contradiction to  the idea behind the Huyghens Principle that each point of a wavefront acts as a secondary source
emitting signal to all space directions; in other words, it assumes, at least in principle, that the signal at a point of a wavefront is made of contributions from all points of previous wavefronts.  This idea could be appropriate for a description of light as a continuous wave manifestation, but not as a discrete one.\\
In contradistinction, the constraints (\ref{lcone}) and (\ref{RdR}), together, imply from the start that a point on a wavefront propagates, on its light-cone generator, independently of all the other wavefront points. Each point of a wavefront, therefore, can be treated as an independent object by itself. It will be shown in section IV that each point of a wavefront is created and annihilated (emitted and absorbed) not in a continuous way as usually thought in classical physics but in a discrete way, like a photon in quantum physics. It is so justified the naming  of a CLASSICAL PHOTON to each point of an electromagnetic wavefront. One can associate to a classical photon the idea of a classical particle of null mass and dimensions. A classical photon is {\it related} to the intersection of (\ref{LWS}),  with a light-cone generator.  The point is that (\ref{LWS}) is a solution of (\ref{BoxA}) which  describes a wave propagating on a light-cone. The appropriate description of a point propagating along a light-cone generator is done by an equation written in terms of the $\nabla$ operator defined below. This is being presented elsewhere \cite{CPMF}, and will not be discussed here. \\

The simultaneous imposition of (\ref{lcone}) and (\ref{RdR}) corresponds then to an EXTENDED CAUSALITY concept applied to massless objects; it is readily extensible
to massive objects too \cite{hep-th/9505169}. It is appropriate for
descriptions of particle-like fields with discrete interactions, that is,
localized and propagating like a particle. Usually field theories are based on the local causality, but it is possible to build  a theory based on
this extended causality \cite{CPMF}.\\
Armed with these concepts of extend-causality and of classical photons one can present another physical
interpretation of the above two Lienard-Wiechert solutions. At the event x there are
two classical photons. One, that was emitted by the electron current $J$, at
$z(\tau_{ret})$ with $x^{0}>z^{0}(\tau_{ret})$, and is moving in the K
generator of the x-light-cone, $K^{\mu}:=(K^{0},{\vec K}).$  $J$ is its
source. The other one, moving on a ${\bar K}$-generator, ${\bar
K}^{\mu}:=(K^{0},-{\vec K}),$ will be absorbed by $J$ at $z(\tau_{adv}),$ with
$x^{0}<z^{0}(\tau_{adv}).$ $J$ is its sink. See figure 2. They are both
retarded solutions and correspond, respectively, to the creation and
destruction of a classical photon. Exactly this: creation and destruction of
particles in classical physics! This interpretation is only allowed with these concepts of extended causality and of classical photon; it is not possible with the continuous wave solutions. It will be instrumental for a clear understanding of how those one-century old problems of CED are all worked out.\\ 

\begin{section}
{ ENERGY TENSOR AND INTEGRABILITY}
\end{section}

When taking derivatives of $A(x)$ the restriction (\ref{RdR}),
or equivalently, $K_{\mu}=-\frac{\partial \tau_{ret}}{\partial x^{\mu}}$ must be considered. This can turn, for the untrained, a trivial calculation into a mess. The best and more fruitful approach, in my opinion, is to take $x$ and $\tau_{ret}$ as 5 independent parameters, and absorb the restriction (\ref{RdR}) in the definition of a new derivative operator $\nabla,$
replacing the usual one:
\begin{equation}
\label{nabla}
\frac{\partial}{\partial
x^{\mu}}\Rightarrow\nabla_{\mu}:=\frac{\partial}{\partial
x^{\mu}}+\frac{\partial\tau}{\partial
x^{\mu}}\frac{\partial}{\partial\tau}=\frac{\partial}{\partial
x^{\mu}}-K_{\mu}\frac{\partial}{\partial\tau},
\end{equation}
or $\nabla_{\mu}:={\partial} _{\mu} -K_{\mu}\partial_{\tau},$ in a shorter
notation. Therefore, $\partial_{\mu}A(x)$, with the explicit restriction ${\big|}_{{\tau}_{ret}}$
 is equivalent to $\nabla_{\mu}A(x)$ without any restriction.
\begin{equation}
\label{nablaa}
\partial_{\mu}A(x){\big|}_{{\tau}_{ret}}=\nabla_{\mu}A(x,\tau)
\end{equation}
  This restriction, ${\big|}_{{\tau}_{ret}},$ is, therefore, implicitly assumed everywhere in this paper except when otherwise is clearly stated, as for example  in the following section when the points $\tau_{ret}\pm d\tau$ are considered. The use of $\nabla$, as defined in (\ref{nablaa}), simplifies the notation, as it is not necessary anymore to carry this restriction ${\big|}_{{\tau}_{ret}}.$\\

The geometric meaning of $\nabla$ is quite clear; it is the
derivative allowed by the restrictions (\ref{lcone}) and (\ref{RdR}), that is, displacements along the K light-cone generator only.
One could complete the geometric picture seeing the operator $\nabla$ as  a kind of ``covariant derivative" with the connections of a new spacetime geometry \cite{hep-th/9505169} that would give a description equivalent to the old Minkowski spacetime plus generalizations of the constraints (\ref{lcone}) and (\ref{RdR}). This would correspond to a complete  geometrization of the extended causality concept.\\ 
It is funny that although the standard view of CED uses the concept of local causality (that is, only the constraint (\ref{lcone})) for interpreting the Lienard-Wiechert solutions, it actually does all further calculation (the Maxwell stress tensor, for example ) according to the rules of the extended causality concept. In other words, the electromagnetic field obtained from (\ref{LWS}) are the variations of (\ref{LWS}) along a lightcone generator.\\

Therefore, 
\begin{equation}
\label{nablaA}
\nabla_{\mu}A^{\nu}=\nabla_{\mu}\frac{V^{\nu}}{\rho}=-\frac{K_{\mu}\hbox{\Large
a}^{\nu}}{\rho}-\frac{V^{\nu}}{\rho^{2}}\nabla_{\mu}\rho=-K_{\mu}\frac{\hbox{\Large a}^{\nu}}{\rho}-\frac{V^{\nu}(K_{\mu}E-V_{\mu})}{\rho^{2}},
\end{equation}
with
\begin{equation}
\label{E}
E=1+\hbox{\Large a}.R=1+\rho\hbox{\Large a}_{K},
\end{equation}
as $\nabla_{\mu}V^{\nu}=-K_{\mu}\hbox{\Large a}^{\nu}$ and
\begin{equation}
\label{nablarho}
\nabla_{\mu}\rho=K_{\mu}E-V_{\mu},
\end{equation}
 where $\hbox{\Large a}_{K}:=\hbox{\Large a}.K.$ For notation simplicity 
$[A,B]$ stands for  $\;[A_{\mu},B_{\nu}]:=A_{\mu}B_{\nu}-B_{\mu}A_{\nu}\;$
and (A,B)  for $(A_{\mu},B_{\nu}):=A_{\mu}B_{\nu}+A_{\nu}B_{\mu}.$
Observe that the Lorentz gauge condition is automatically satisfied
\begin{equation}
\label{Lgc}
\nabla.A=-\frac{\rho\hbox{\Large a}_K+V.\nabla\rho}{\rho^{2}}=0,
\end{equation}
as $V.K=-1$, $V^{2}=-1$, and $V.\nabla\rho=1-E=-\rho\hbox{\Large a}_K$.\\
The Maxwell field ${F_{\mu\nu}}:=\nabla_{\nu}A_{\mu}-\nabla_{\mu}A_{\nu}$, is
found to be
\begin{equation}
\label{F}
{F}=\frac{1}{\rho^{2}}[K,W],
\end{equation}
with
\begin{equation}
\label{W}
W^{\mu}=\rho\hbox{\Large a}^{\mu}+EV^{\mu}.
\end{equation}
The electron self-field energy-momentum tensor,
$4\pi\Theta=F.F-\frac{\eta}{4}F^{2}$, is
\begin{equation}
\label{t}
4\pi\rho^{4}\Theta^{\mu\nu}=[K^{\mu},W^{\alpha}][K_{\alpha},W^{\nu}]-\frac{{\eta}^{\mu\nu}}{4}[K^{\alpha},W^{\beta}][K_{\beta},W_{\alpha}],
\end{equation}
or in an expanded expression
\begin{equation}
\label{t'}
4\pi\rho^{4}\Theta=(K,W)+KKW^{2}+WWK^{2}+\frac{\eta}{2}(1-K^{2}W^{2}),
\end{equation}
as $K.W=-1$. The use of rather compact expressions like (\ref{t}) instead of
(\ref{t'}) is preferred because besides being compact they will make easier the calculation of the zero-distance limit in the following sections. With $W^{2}=\rho^{2}\hbox{\Large
a}^{2}-E^{2}=\rho^{2}\hbox{\Large a}^{2}-(1+\rho\hbox{\Large a}_K)^{2},$
$\Theta$ may be written, according to its powers of $\rho,$ as
$\Theta=\Theta_{2}+\Theta_{3}+\Theta_{4}.$
If the $K^{2}$-terms are neglected then
\begin{equation}
\label{t2e}
4\pi\rho^{2}\Theta_{2}{\big|}_{K^{2}=0}=-KK(\hbox{\Large
a}^{2}-{\hbox{\Large a}_{K}}^{2}),
\end{equation}
\begin{equation}
\label{t3e}
4\pi\rho^{3}\Theta_{3}{\big|}_{K^{2}=0}=2KK\hbox{\Large a}_K -
(K,\hbox{\Large a}+V\hbox{\Large a}_K),
\end{equation}
\begin{equation}
\label{t4e}
4\pi\rho^{4}\Theta_{4}{\big|}_{K^{2}=0}=KK -(K,V) -\frac{\eta}{2},
\end{equation}
which are the usual expressions that one finds, for example in
\cite{Rohrlich,Jackson,Parrot,Teitelboim,Rowe,Lozada}.
Observe that
\begin{equation}
\label{C}
K.\Theta_{2}{\big|}_{K^{2}=0}=0,
\end{equation}
which is important in the identification of $\Theta_{2}$ with the radiated
\cite{Teitelboim} part of $\Theta$, and that
\begin{equation}
\label{C1} K.\Theta_{3}{\big|}_{K^{2}=0}=0.
\end{equation}
The presence of non-integrable singularities in the electron self-field energy
tensor is a major problem.
$\Theta_{2}{\big|}_{K^{2}=0}$, although singular at $\rho=0$, is nonetheless
integrable. By that it is meant that it produces a finite  flux through a
spacelike hypersurface $\sigma$ of normal $n$, that is,  $\int
d^{3}\sigma\Theta_{2}.n$ exists \cite{Rowe}, while
$\Theta_{3}{\big|}_{K^{2}=0}$ and $\Theta_{4}{\big|}_{K^{2}=0}$ are not
integrable; they generate, respectively, the problematic Schott term in the LDE
and a divergent term, the electron bound 4-momentum \cite{Teitelboim}, which
includes the so called electron self-energy.
Previous attempts, based on distribution theory, for  taming these
singularities have relied on modifications of the Maxwell
theory with  addition of extra terms to $\Theta{\bigg|}_{K^{2}=0}$ on the
electron world-line (see for example the reviews
\cite{Teitelboim,Rowe,Lozada}). They redefine $\Theta_{3}{\bigg|}_{K^{2}=0}$
and $\Theta_{4}{\bigg|}_{K^{2}=0}$ at the electron world-line in order to make
them integrable without changing them at $\rho>0,$  so to preserve the standard
results of Classical Electrodynamics. But this is always an ad hoc introduction
of something strange to the theory. Another unsatisfactory aspect of this
procedure is that it regularizes the above integral but leaves an unexplained
and unphysical discontinuity in the flux of 4-momentum, $\int
dx^{4}\Theta^{\mu\nu}\nabla_{\nu}\rho\;\delta(\rho-\varepsilon),$ through a
 cylindrical hypersurface $\rho=\varepsilon=const$ enclosing  the charge world-line.
\noindent It is particularly interesting that, as it will be shown in the sequence, instead of adding anything one should actually not drop out the null $K^2$-terms. Their
contribution (not null, in an appropriate limit) cancel the infinities. The
same problem happens in the derivations of the electron equation of motion from
these incomplete expressions of $\Theta.$  The Schott term in the Lorentz-Dirac
equation is a consequence; it does not
appear in the equation when the full expression of $\Theta$ is correctly used. The point is that K and $\Theta$ are defined only for $\rho>0$. $K^{2}=0$ is also true only for $\rho>0.$ Everybody in the literature uses not the complete expression (\ref{t'}) for $\Theta$, but the shorter $\Theta{\big|}_{K^{2}=0}$-expressions when considering the limit of $\rho$ tending to zero. Therefore, there is a generalized use of an implicit assumption that $K^{2}$ remains null at the limit $\rho=0.$ This is false, as it is shown in the next section, and compromises all the literature results.\\

\begin{section}
{ THE ZERO-DISTANCE LIMIT}
\end{section}
$\Theta$ is an explicit function of K and $\rho.$
K is defined only for $\rho>0,$  $K:=\frac{R}{\rho},$  and so is also $K^{2}=0.$ At the limiting point $\rho=0$ they produce a $(\frac{0}{0})-$type of
indeterminacy, as R necessarily tends also to zero: ($R\rightarrow0$) or $x\rightarrow z(\tau_{ret}),$ along the lightcone generator $K^{\mu}.$
 By force of the constraints (\ref{lcone}) and (\ref{RdR}), as x and
$z(\tau_{ret})$ must remain on a same straight-line, the lightcone-generator K,
 the limit $\rho\rightarrow0$  necessarily implies also on $x^{\mu}\rightarrow
z(\tau_{ret})^{\mu}$ or $R^{\mu}\rightarrow 0.$\\
The $(\frac{0}{0})-$type of
indeterminacy of $K=\frac{R}{\rho}$ at $z(\tau_{ret})$, can be evaluated at  neighboring points $\tau=\tau_{ret}\pm d\tau$ by the L'H\^opital's rule and
$\frac{\partial}{\partial\tau}$ (see figure 3).
This application of the L'H\^opital's rule corresponds then to finding two
simultaneous limits: $\rho\rightarrow0$ and $\tau\rightarrow\tau_{ret}$.\\ As
\begin{equation}
\label{rodot}
\partial_{\tau}\rho\equiv\stackrel{.}{\rho}=-(1+\hbox{\Large a}.R)
\end{equation}
 and
\begin{equation}
\label{rdot}
\stackrel{.}{R}=-V,
\end{equation}
 then
\begin{equation}
\label{limitK}
\lim_{\rho\to0\atop \tau\to\tau_{ret}} K{\Big|}_{ R^{2}=0\atop R.dR=0} = V.
\end{equation}
This double limiting process is of course distinct of the single
$(\rho\rightarrow0)$-limit, which cannot avoid the singularity.
For notation simplicity  the use of just $lim_{\rho\rightarrow0}$ will be kept but always  with the implicit meaning of this double limit as indicated in (\ref{limitK}). For example
by
\begin{equation}
\label{limitK2}
\lim_{\rho\to0}K^{2}=-1.
\end{equation}
it is meant
\begin{equation}
\label{limitK2T}
\lim_{\rho\to0\atop \tau\to\tau_{ret}}K^{2}{\Big|}_{ R^{2}=0\atop R.dR=0}=-1.
\end{equation}
This invalidates all previous results in the literature on the CED short distance behaviour because they all have been obtained from $\Theta{\big|}_{K^{2}}$ as it is valid only for $\rho>0$; but then it could not be used in the ($\rho\rightarrow0$)-limit. Besides, the correct limit (\ref{limitK}) has not been used.

This limit (\ref{limitK}) and the geometry behind it require a consistent  physical interpretation that implies on a new connection between classical and quantum physics. The classical electromagnetic interaction  between two point charges as described by the Lienard-Wiechert solution $A(x)$,   comprises the entire light-cone, $R^{2}=0,$ that is, all the space surrounding each charge. But the simultaneous imposition of $R^{2}=0$ and $R.dR=0$ (or $d\tau+K.dx=0$) implies that only the part of $A(x)$ contained in the light-cone generator, K, connecting the two charges must be considered at a time. This is the possible description, in classical physics, of the electromagnetic fundamental interaction: the exchange of a single photon. The light-cone generator is the photon classical trajectory. See figures 2 and 3. Now it is possible to understand the reasons of the $(\frac{0}{0})$-indeterminacy at $\tau=\tau_{ret}.$
In the limit of $\rho\rightarrow0$  at $\tau=\tau_{ret}$ there are  3 distinct velocities: K, the photon 4-velocity, and $V_{1}$ and $V_{2}$, the electron initial and final 4-velocities. The singularity at $\tau_{ret}$ is not associated to any infinity but to an indeterminacy in the tangent of the electron worldline.  At $\tau=\tau_{ret}+d\tau$ there 
is only $V_{2}$, and only $V_{1}$ at $\tau=\tau_{ret}-d\tau.$  In other words, $\tau_{ret}$ is an isolated singular point on the electron world-line; its neighboring points $\tau_{ret}\pm d\tau$ are not singular.  This is in flagrant contradiction to the Classical-Electrodynamics assumption of a continuous emission process, because in this case, all points on the electron world-line would be singular points, like $\tau_{ret}$. This completes the justification for the introduction of the classical photon concept: the part of the electromagnetic interaction contained  in a lightcone generator is independent of the other parts contained in the other lightcone generators and, besides, it is discretely emitted and absorbed. There is a classical photon at $\tau_{ret}$ but there is none at $\tau_{ret}\pm d\tau.$ This picture will receive a further confirmation by the calculation of the energy flux from the charge at $z(\tau_{ret}\pm d\tau)$, in section VII.
It is remarkable that one can find in a classical  (Lienard-Wiechert) solution these traits of the quantum nature of the radiation emission process.
They show a new bridge between classical and quantum field theories: the classical field is an effective representation of the effects of a photon exchange smeared in the charge light-cone. $R^{2}=0$ and $R.dR=0$ establish an extended constraint of causality that retrieves from the smeared-interaction field the interaction one-photon-exchange character.\\ 

\begin{section}
{ Some useful mathematical tools}
\end{section}

To find this double limit of something when $\rho\rightarrow0$ and
$\tau\rightarrow\tau_{ret}$ will be done so many times in this paper that it is
better to do it in a more systematic way.   One wants to find
\begin{equation}
\label{LR}
\lim_{\rho\to0}\frac{N(R,\dots)}{\rho^{n}},
\end{equation}
where $N(R,\dots)$ is a homogeneous function of R,
$N(R,\dots){\big|}_{R=0}=0$. Then, one has to apply the L'H\^opital's rule
consecutively until the indeterminacy is resolved.  As
$\frac{\partial\rho}{\partial\tau}=-(1+\hbox{\Large a}.R)$, the denominator of
(\ref{LR}) at $R=0$ will be different of zero only after the
$n^{th}$-application of the L'H\^opital's rule, and then, its value will be
$(-1)^{n}n!$\\
If $p$ is the smallest integer such that $N(R,\dots)_{p}{\big|}_{R=0}\ne0,$
where $N(R)_{p}:=\frac{d^{p}}{d{\tau}^{p}}N(R,\dots)$, then
\begin{equation}
\label{NR}
\lim_{\rho\to0}\frac{N(R,\dots)}{\rho^{n}}=\cases{\infty,& if $p<n$\cr
              (-1)^{n}{\frac{N(0,\dots)_{p}}{n!}},& if $p=n$\cr
0,& if $p>n$\cr}
\end{equation}
\begin{itemize}
\item Example 1: $\cases{K=\frac{R}{\rho}.& $n=p=1 \Longrightarrow
\lim_{\rho\to0}K=V$\cr K^{2}=\frac{R.\eta.R}{\rho^{2}}.&$ n=p=2 \Longrightarrow
\lim_{\rho\to0}K^{2}=-1.$\cr}$
\item Example 2: $\frac{[K,\hbox{\Large a}]}{\rho}=\frac{[R,\hbox{\Large
a}]}{\rho^{2}}\;\;$ $\;\Longrightarrow \;$ $\;p=1<n=2\Longrightarrow
\lim_{\rho\to0}\frac{[K,\hbox{\Large a}]}{\rho}$ diverges
\item Example 3: $\frac{\hbox{\Large a}_K}{\rho}[K,V]=-\frac{\hbox{\Large
a}.R}{\rho^{3}}[R,V]\Longrightarrow p=4>n=3$ \qquad
$\lim_{\rho\to0}\frac{\hbox{\Large a}_K}{\rho}[K,V]=0$
\item Example 4: $\frac{[K,V]}{\rho^{2}}=\frac{[R,V]}{\rho^{2}}\;\;
\;\Longrightarrow
\;\;p=2<n=3\Longrightarrow\lim_{\rho\to0}\frac{[K,V]}{\rho^{2}}$ diverges
\end{itemize}
Finding these limits for more complex functions can be made easier with two
helpful expressions,
\begin{equation}
\label{NP}
N_{p}=\sum_{a=0}^{p}\pmatrix{p\cr a\cr}A_{p-a}.B_{a}
\end{equation}
and
\begin{equation}
\label{AgBC}
N_{p}=\sum_{a=0}^{p}\sum_{c=0}^{a}\pmatrix{p\cr a\cr}\pmatrix{a\cr
c\cr}A_{p-a}.B_{a-c}.C_{c}
\end{equation}
valid when $N(R)$ has, respectively, the forms $N_{0}=A_{0}.B_{0},$ or
$N_{0}=A_{0}.B_{0}.C_{0},$ where A, B and C represent possibly distinct
functions of R, and the subindices indicate the order of $\frac{d}{d\tau}$. For
example: $A_{0}=A$;  $A_{1}=\partial_{\tau} A$; $A_{2}=\partial_{\tau}^{2} A,$
and so on.
So, for using (\ref{NR}-\ref{AgBC}), one just has to find the
$\tau$-derivatives of A, B and C that produce the first non- null term at the
point limit of $R\rightarrow0.\;\;$\\
Consecutive derivatives of products of complex functions can become unwieldy. So it is worthy to introduce the concept of ``$\tau$-order" of a function, meaning the lowest order of the $\tau$-derivative of a function that produces a non-null result at the limiting point $R=0$. Let ${\cal O}[f(x)]$ represent the $\tau$-order" of f(x). So, for example, from (\ref{rodot}) and (\ref{rdot}) one sees that
\begin{equation}
\label{tor}
{\cal O}[R]=1,
\end{equation}
\begin{equation}
\label{torho}
{\cal O}[\rho]=1,
\end{equation}
As $\partial_{\tau}(\hbox{\Large a}.R)=-{\dot{\hbox{\Large a}}}.R$ and
$\partial^{2}_{\tau}(\hbox{\Large a}.R)=-{\ddot{\hbox{\Large
a}}}.R-{\dot{\hbox{\Large a}}}.V=-{\ddot{\hbox{\Large a}}}.R+\hbox{\Large
a}^{2}=\hbox{\Large a}^{2}+{\cal O}(R),$ then
\begin{equation}
\label{toar}
{\cal O}[\hbox{\Large a}.R]=2,
\end{equation}
For finding the $N_{p}$ of (\ref{NP}) and of (\ref{AgBC}) it is then necessary
to consider only the terms with the lowest $\tau$-order on each factor. Some
combinations of terms have derivatives that cancel parts of each other
resulting in a higher $\tau$-order term. For example,
$$\partial_{\tau}(R^{2}+\rho^{2})=+2\rho-2\rho E=-2\rho\hbox{\Large a}.R$$
$$\partial^{2}_{\tau}(R^{2}+\rho^{2})=2E\hbox{\Large
a}.R-2\rho{\dot{\hbox{\Large a}}}.R= 2(\hbox{\Large a}.R-\rho{\dot{\hbox{\Large
a}}}.R)+{\cal O}(R^{4}),$$
$$\partial^{3}_{\tau}(R^{2}+\rho^{2})=2({\dot{\hbox{\Large
a}}}.R+E{\dot{\hbox{\Large a}}}.R-\rho\hbox{\Large a}^{2})+{\cal
O}(R^{3})=4{\dot{\hbox{\Large a}}}.R-2\rho\hbox{\Large a}^{2}+{\cal
O}(R^{3}),$$
$$\partial^{4}_{\tau}(R^{2}+\rho^{2})=4\hbox{\Large a}^{2}+2\hbox{\Large
a}^{2}+{\cal O}(R^{2})=6\hbox{\Large a}^{2}+{\cal O}(R^{2}).$$
So, $${\cal O}[R^{2}+\rho^{2}]=4$$ although $${\cal O}[R^{2}]={\cal
O}[\rho^{2}]=2.$$
Observe that one has to care only with the lowest $\tau$-order terms as the
other ones, grouped in ${\cal O}(R)$, will not survive the limit
$R\rightarrow0$. Also, it is not necessary to write the $\tau$-derivatives of
factors that will not reduce its $\tau$-order. For example in
$$\partial_{\tau}(RV+{\cal O}(R^{2}))=-VV+{\cal O}(R^{2}),$$  the term
$R\hbox{\Large a}$ was absorbed in ${\cal O}(R^{2})$. It is avoided, in this way, taking unnecessary derivatives and the writing of long expressions with terms that will not contribute to the final result.\\

An important property of  ${\cal
O}[f(x)]:$
\begin{equation}
\label{ptor}
{\cal O}[ABC]={\cal O}[A]+{\cal O}[B]+{\cal O}[C],
\end{equation}
so that, p as defined by (\ref{NR}) and (\ref{AgBC}), is
\begin{equation}
\label{p}
p={\cal O}[ABC]={\cal O}[A]+{\cal O}[B]+{\cal O}[C].
\end{equation}
\begin{section}
{ FLUXES AND EQUATION OF MOTION}
\end{section}

\noindent The motion of a classical electron \cite{Rohrlich,Jackson,Parrot} is
described by the Lorentz-Dirac equation,
\begin{equation}
\label{LDE}
m\hbox{\Large a} = F_{ext}.V+\frac{2}{3}(\stackrel{.}{\hbox{\Large
a}}-\hbox{\Large a}^{2}V),
\end{equation}
where  m is the electron mass and $F_{ext}$ is an external electromagnetic
field.
The presence of the Schott term, $\frac{2}{3}\!\stackrel{.}{a}$, is  the
cause  of  all of its pathological features, like microscopic non-causality,
runaway solutions, preacceleration, and other bizarre effects \cite{Eliezer}.
On the other hand, its  presence is apparently necessary for the
energy-momentum  conservation; without it it would be required a contradictory
null radiance for an accelerated charge, as $a.V=0,\quad V.F_{ext}.V=0$ and  $\stackrel{.}{a}.V+a^{2}=0$. This
makes of the Lorentz-Dirac equation the greatest paradox of classical field
theory as it cannot simultaneously preserve both the causality and the energy
conservation \cite{Rohrlich,Jackson,Parrot}.\\
The Lorentz-Dirac equation can be obtained from energy-momentum conservation,
which leads to
\begin{eqnarray}
\lefteqn{
 \int_{\tau_{1}}^{\tau_{2}}d\tau(F_{ext}^{\mu\nu}V_{\nu}- m\hbox{\Large a}^{\mu})=\lim_{{\varepsilon_{1}\to0}\atop{{\varepsilon_{2}\to\infty}}}\int_{\cal V}dx^{4}\nabla_{\nu}\Theta^{\mu\nu}\theta(\varepsilon_{2},\rho,\varepsilon_{1})\theta(\tau_{2},\tau,\tau_{1})=}
\hspace{30ex}\nonumber\\
& &                     =\lim_{{\varepsilon_{1}\to0}\atop{{\varepsilon_{2}\to\infty}}}\int_{\cal V}  dx^{4}\Theta^{\mu\nu}{\Big\lbrace}\nabla_{\nu}\rho\;{\Big (}\theta(\rho-\varepsilon_{1})\delta(\varepsilon_{2}-\rho)
-\delta(\rho-\varepsilon_{1})\theta(\varepsilon_{2}-\rho){\Big)\theta(\tau_{2},\tau,\tau_{1}}{\Big)}+ \nonumber\\ 
& & 
+\nabla_{\nu}\tau{\Big (}\delta(\tau_{2}-\tau)\theta(\tau-\tau_{1})-\theta(\tau_{2}-\tau)\delta(\tau-\tau_{1}){\Big )}\theta(\varepsilon_{2},\rho,\varepsilon_{1}){\Big\rbrace},\label{eem}
\end{eqnarray}
where $\theta(a_{2},x,a_{1})=\theta(a_{2}-x)\theta(x-a_{1})$, the product of two Heaviside functions,  and $\;\tau_{2},\;\tau_{1},\;\varepsilon_{2}\;\hbox{and}\;\varepsilon_{1}$ are constants, with $\tau_{2}>\tau_{1}$ and $\varepsilon_{2}>\varepsilon_{1}$.
$\theta(\tau_{2},\tau,\tau_{1})=\theta(\tau_{2}-\tau)\theta(\tau-\tau_{1})$ defines the spacetime region   between the two light-cones of vertices at $\tau_{2}$ and $\tau_{1}.$ The product of these 4 Heaviside functions defines the closed boundary of an hypervolume that is inside the integration domain ${\cal V}.$ The passage from the first to the second and third lines of (\ref{eem}) involves integration by parts, the divergence theorem and the use of
\begin{equation}
\label{V}
\Theta^{\mu\nu}\theta(\varepsilon_{2},\rho,\varepsilon_{1})\theta(\tau_{2},\tau,\tau_{1}){\Big|_{\partial{\cal V}}}=0.
\end{equation}
 $\nabla_{\nu}\Theta^{\mu\nu}=0$ in the hypervolume, for $\varepsilon_{1}>0$ assures that the integral in the RHS of the first line of (\ref{eem}) is null for any $\varepsilon_{1}>0,$ but not, as it will be shown now, in the limit when $\varepsilon_{1}$ tends to zero. This approach is equivalent to one where $\Theta^{\mu\nu}$ is treated as a distribution \cite{Rowe,Lozada}. Both are equally rigourous and give the same results, but this one is simpler as it dispenses the use of a compact test function, which is replaced by $\theta(\varepsilon_{2}-\rho),$ in its role of allowing a compact domain of integration. No infinity appears in this approach and so it is not necessary to consider the distribution character of $\Theta^{\mu\nu}.$
The terms in the second and third lines of (\ref{eem}) are  the fluxes of energy-momentum through the respective hypersurfaces $\rho=\varepsilon_{1},$  $\rho=\varepsilon_{2},$ $\tau=\tau_{2}$ and $\tau=\tau_{1}$. They are well known in the literature\cite{Rohrlich,Marx}, for $\varepsilon_{1}>0.$ The flux on a Bhabha tube $\rho=\varepsilon>0$ is given by 
\begin{equation}
\label{fbt}
\Phi_{1}(\varepsilon)^{\mu}=\int dx^{4}\Theta^{\mu\nu}\nabla_{\nu}\rho\;\delta(\varepsilon-\rho)\theta(\tau_{2},\tau,\tau_{1}).
\end{equation}
With 
\begin{equation}
\label{nablarho2}
W.\nabla\rho\equiv0,
\end{equation}
\begin{equation}
\label{KW}
K.W=-1,
\end{equation}
\begin{equation}
\label{Knro}
K.\nabla\rho=1+K^{2}E,
\end{equation}
and $K^{2}=0$, as $\varepsilon>0$, in (\ref{t}) one has
\begin{equation}
4\pi\rho^{4}\Theta.\nabla\rho=W+K(W^{2}+\frac{E}{2})-\frac{1}{2}V=\rho\hbox{\Large a}+V(\rho\hbox{\Large a}_{K}+\frac{1}{2})+K{\Big(}\rho^{2}(\hbox{\Large a}^{2}-\hbox{\Large a}_{K}^{2})-\frac{1}{2}(1+3\rho\hbox{\Large a}_{K}){\Big)}.
\end{equation}
Now using retarded coordinates \cite{Teitelboim,Rowe,Synge} where $d^{4}x=d\tau \rho^{2}d\rho d^{2}\Omega$, and 
\begin{equation}
\frac{1}{4\pi}\int d^{2}\Omega K^{\alpha}=V^{\alpha},
\end{equation}
\begin{equation}
\frac{1}{4\pi}\int d^{2}\Omega K^{\alpha}K^{\beta}=\frac{1}{3}\Delta^{\alpha\beta}+V^{\alpha}V^{\beta},
\end{equation}
\begin{equation}
\frac{1}{4\pi}\int d^{2}\Omega K^{\alpha}K^{\beta}K^{\gamma}=\Delta^{(\alpha\beta}V^{\gamma)}+V^{\alpha}V^{\beta}V^{\gamma},
\end{equation}
where $\Delta=\eta+VV,$ and the parenthesis on the superscripts mean total symmetrization, one has for (\ref{fbt}) 
\begin{equation}
\Phi_{1}(\varepsilon)^{\mu}=\int_{\tau_{1}}^{\tau_{2}}d\tau(\frac{2}{3}{\hbox{\Large a}}^{2}V^{\mu}-\frac{\hbox{\Large a}^{\mu}}{2\varepsilon}).
\end{equation}

The total flux on the sections $0<\varepsilon_{1}<\rho<\varepsilon_{2}$ of the lightcones $\tau=\tau_{2}$ and $\tau=\tau_{1}$ is given by 
\begin{eqnarray}
\lefteqn{\Phi_{2}(\varepsilon_{2})^{\mu}-\Phi_{2}(\varepsilon_{1})^{\mu}=-\int dx^{4}\Theta^{\mu\nu}K_{\nu}\;\theta(\varepsilon_{2},\rho,\varepsilon_{1}){\Big(}\delta(\tau_{2}-\tau)-\delta(\tau-\tau_{1}){\Big)}=}
\hspace{15ex}\nonumber\\
& &                     
=\frac{1}{2}{\Big(}V(\tau_{2})^{\mu}-V(\tau_{1})^{\mu}{\Big)}\;(\frac{1}{\varepsilon_{2}}-\frac{1}{\varepsilon_{1}})=\frac{1}{2}\int_{\tau_{1}}^{\tau_{2}}d\tau\hbox{\Large a}^{\mu}(\frac{1}{\varepsilon_{2}}-\frac{1}{\varepsilon_{1}}),\label{flc}
\end{eqnarray}
as $\nabla_{\nu}\tau=-K_{\nu}$ and $4\pi\rho^{4}\Theta.K=\frac{1}{2}K,$ for $\rho>0.$
In the notation used in (\ref{flc}), $\Phi_{2}(\varepsilon_{2})$ and 
$\Phi_{2}(\varepsilon_{1})$ are, respectively, the upper and the lower limit of the $\rho$-integration in (\ref{flc}): \begin{equation}
\Phi_{2}(\varepsilon_{2})=\frac{1}{2}\int_{\tau_{1}}^{\tau_{2}}d\tau\frac{\hbox{\Large a}}{\varepsilon_{2}}
\end{equation}
 and 
\begin{equation}
\Phi_{2}(\varepsilon_{1})=\frac{1}{2}\int_{\tau_{1}}^{\tau_{2}}d\tau\frac{\hbox{\Large a}}{\varepsilon_{1}}.
\end{equation}
One sees then from (\ref{fbt}) and (\ref{flc}) that the integral in the RHS of the first line of (\ref{eem}) is equal to
\begin{equation} \int dx^{4}\nabla_{\nu}\Theta^{\mu\nu}\theta(\varepsilon_{2},\rho,\varepsilon_{1})\theta(\tau_{2},\tau,\tau_{1})=\Phi_{1}(\varepsilon_{2})-\Phi_{1}(\varepsilon_{1})+\Phi_{2}(\varepsilon_{2})-\Phi_{2}(\varepsilon_{1})
\end{equation}
 and that this (the RHS) is identically null, for any $\varepsilon_{1}>0,$ which is in agreement with $\nabla_{\nu}\Theta^{\mu\nu}$ being null at $\rho>0.$ These results can be extended to $\varepsilon_{2}\rightarrow\infty$ but not to $\varepsilon_{1}\rightarrow0$ because of the explicit dependence of $\Theta^{\mu\nu}$ on the null 4-vector K, which is defined only for $\rho>0.$ At $\rho=0$ its definition ($K=\frac{R}{\rho})$ gives a $(\frac{0}{0})-$indeterminacy. In the literature it is implicitly assumed that K remains a null 4-vector at the limiting $\rho=0.$ Besides not being correct, as it has been shown, this produces a diverging flux (the self energy problem) and the controversial Schott term in the Lorentz-Dirac equation. 

In the $(\varepsilon_{2}\rightarrow\infty)$-limit one has $\Phi_{2}(\varepsilon_{2}=\infty)=0$ and $\Phi_{1}(\varepsilon_{2}=\infty)=\int_{\tau_{1}}^{\tau_{2}}d\tau\frac{2}{3}{\hbox{\Large a}}^{2}V^{\mu}.$ Therefore,  with (\ref{fbt}) and (\ref{flc}), eq. (\ref{eem}) can be written as
\begin{equation}
\label{ef}
 \int_{\tau_{1}}^{\tau_{2}}d\tau(F_{ext}^{\mu\nu}V_{\nu}- m\hbox{\Large a}^{\mu})=-\lim_{\varepsilon_{1}\to0}{\big\{}\Phi_{2}(\varepsilon_{1})^{\mu} +\Phi_{1}(\varepsilon_{1})^{\mu}{\big\}}+\int_{\tau_{1}}^{\tau_{2}}d\tau\frac{2}{3}{\hbox{\Large a}}^{2}V^{\mu}.
\end{equation}

(\ref{NR}-\ref{AgBC}) will now be used for finding the $(\varepsilon_{1}\rightarrow0)$-limit of $\Phi_{1}(\varepsilon_{1})=\int^{\tau_{2}}_{\tau_{1}}d\tau \rho^{2}d\rho\;d^{2}\Omega\;\Theta.\nabla\rho\;\delta(\rho-\varepsilon_{1}).$  But in (\ref{t}), the definition of $\Theta$, the second term is the trace of the
first one and so one just has to consider this last one because the behaviour
of its trace under this limiting process can then easily be inferred. So, as
$K=\frac{R}{\rho},$ and $\nabla\rho=(KE-V)$ one has schematically, for the
first term of (\ref{t}) in $\rho^{2}\Theta.\nabla\rho$,
\begin{equation}
\label{10}
\lim_{\rho\to0}\frac{N(R,\dots)}{\rho^{n}}=\lim_{\rho\to0}\frac{\rho^{2}[K,W].[K,W].(KE-V)}{\rho^{4}}=\lim_{\rho\to0}\frac{[R,W].[R,W].(RE-V\rho)}{\rho^{5}}
\end{equation}
Then, from the comparison with (\ref{LR}) and (\ref{AgBC}),
\begin{equation}
A_{0}=B_{0}=[R,W]=[R,\hbox{\Large a}\rho+VE]=[R,\hbox{\Large a}\rho+V]+{\cal
O}(R^{3})
\end{equation}
$$ A_{1}=B_{1}=[-V,\rho\hbox{\Large a}+V]+[R,-\hbox{\Large a}E+\hbox{\Large
a}]+{\cal O}(R^{2})=-[V,\rho\hbox{\Large a}]+{\cal O}(R^{2});$$
$$ A_{2}=B_{2}=-[\hbox{\Large a},V]+{\cal O}(R);$$
\begin{equation}
C_{0}=RE-V\rho=R-V\rho+{\cal O}(R^{3}),
\end{equation}
$$ C_{1}=-V-\hbox{\Large a}\rho+VE+{\cal O}(R^{2})=-\hbox{\Large a}\rho+{\cal
O}(R^{2}),$$
$$ C_{2}=\hbox{\Large a}+{\cal O}(R).$$
Therefore, for producing a possibly non null $N_{p}$, according to
(\ref{AgBC}),  $a, c$ and $p$  must be given by  $$c=2,$$
$$p-a=a-c=2\Longrightarrow p=6>n=5.$$
Or in a shorter way, $${\cal O}[[R,W]]=2,$$ $${\cal O}[RE-V\rho]=2,$$
and then, using (\ref{p}), $$p=2{\cal O}[[R,W]]+{\cal O}[RE-V\rho]=6>n=5.$$
Therefore, 
\begin{equation}
\label{fpe}
\lim_{\varepsilon_{1}\to0}\Phi_{1}(\varepsilon_{1})=0.
\end{equation}
The flux of energy and momentum of the electron self-field through the $(\rho=\varepsilon_{1})$-hypersurface in (\ref{eem}) is null at $\varepsilon_{1}=0.$  This is a new result, a consequence of (\ref{limitK}). In the standard approach, with the uncomplete expressions of $\Theta^{\mu\nu}$,  the contribution from this term produces the problematic Schott term and a diverging expression, the electron bound-momentum which requires  mass renormalization \cite{Teitelboim1}. 
If one had used in (\ref{fpe}) the $K^{2}$-terms expurgated energy tensor, which is the one used in the literature, one would have found an infinity on its RHS, even using (\ref{limitK}). The $K^{2}$-terms in (\ref{t}) cancel the infinities.\\
For the  evaluation of $\lim_{\varepsilon_{1}\to0}\Phi_{2}(\varepsilon_{1})$ one finds from  (\ref{t'}) and $K.W=-1$ that 
\begin{equation}
\label{eflc2}
4\pi\rho^{2}\Theta^{\mu\nu}K_{\nu}=\frac{K^{\mu}}{2\rho^{2}}(1-K^{2}W^{2})=\frac{K^{\mu}}{2}{\big\{}\frac{1+K^{2}}{\rho^{2}}-2\frac{K^{2}\hbox{\Large a}.K}{\rho}-K^{2}(\hbox{\Large a}^{2}-{\hbox{\Large a}_K}^{2}){\big\}},
\end{equation}
and 
\begin{equation}
\label{2eflc}
4\pi\int d\rho\rho^{2}\Theta^{\mu\nu}K_{\nu}=-\frac{K^{\mu}}{2}\{\frac{1+K^{2}}{\rho}+2K^{2}\hbox{\Large a}.K\;ln\rho+K^{2}(\hbox{\Large a}^{2}+{\hbox{\Large a}_K}^{2})\rho\}.
\end{equation}
But, again from comparison with (\ref{LR}-\ref{AgBC}),
\begin{equation}
\label{3eflc} 
\lim_{\rho\to0}\frac{K^{\mu}(1+K^{2})}{\rho}=\frac{R^{\mu}(\rho^{2}+R^{2})}{\rho^{4}}=0,
\end{equation}
as ${\cal O}[R]+{\cal O}[\rho^{2}+R^{2}]=1+4>4.$\\Also, as
\begin{equation}
\label{O}
{\cal O}[K^{\mu}K^{2}\hbox{\Large a}.K]=1,
\end{equation}
one can say  from (\ref{torho}) that 
\begin{equation}
\label{p1}
\lim_{\rho\to0}K^{\mu}K^{2}\hbox{\Large a}.K\sim \lim_{\rho\to0}\rho,
\end{equation}
that is, $K^{\mu}K^{2}\hbox{\Large a}.K$ tends to zero with $\rho$ as fast as $\rho$. So,
\begin{equation}
\label{4eflc}
\lim_{\rho\to0}K^{\mu}K^{2}\hbox{\Large a}.K\;ln\rho= \lim_{\rho\to0}\rho\;ln\rho=0.
\end{equation}
The last term of (\ref{2eflc}) is also null in the $(\rho\rightarrow0)$-limit,  and so one can conclude that
\begin{equation}
\label{5eflc}
\lim_{\rho\to0}\Phi_{2}(\varepsilon_{1})=0.
\end{equation}
The meaning of (\ref{fpe}) and (\ref{5eflc}): at $\rho=0$ and $\tau=\tau_{ret}\pm d\tau$ there is only the electron! No self-field, no photon! The flux from the charge is zero for $\tau_{ret}\pm d\tau$ and, of course, non-zero at $z(\tau_{ret}).$ This confirms the picture of a discrete radiation process. It takes the limiting $\rho\rightarrow0$ to be seen because at $\rho>0$ it is masqueraded by the field average character, as it will be discussed later. This is in contradiction to the Gauss's law! It requires a revision of its physical meaning  and of the Maxwell-Faraday concept of field, which will be done on section IX. First one should discuss the issue of the electron equation of motion which will make more evident the inadequacy of the picture of a continuous interaction in a short distance scale.\\

\begin{section}
{ AN EFFECTIVE EQUATION OF MOTION}
\end{section}

 With  (\ref{fpe})  and (\ref{5eflc}) in (\ref{ef}) one could  write the electron equation of motion as
 \begin{equation}
\label{fld}
m\hbox{\Large a}^{\mu}-F^{\mu\nu}_{ext}V_{\nu}=-\frac{2}{3}\hbox{\Large
a}^{2}V^{\mu},
\end{equation}
but it is well known that this could not be a correct equation because it is
not self-consistent: its LHS is orthogonal to V,
\begin{equation}
\label{aveo}
m\hbox{\Large a}.V=0\qquad\hbox{and}\qquad V.F_{ext}.V=0,
\end{equation}
 while its RHS is not,
\begin{equation}
-\frac{2}{3}\hbox{\Large a}^{2}V.V=\frac{2}{3}\hbox{\Large a}^{2}.
\end{equation}
 This seems to be paradoxical until one has a clearer idea of what is
happening. One must return  to equation (\ref{eem}), where there is
a subtle and very important distinction between its LHS and its RHS. Its LHS is entirely determined by the electron instantaneous position, $z(\tau),$ while its RHS is determined by the sum of contributions from the electron self-field at all points. The equation of motion is the mathematical description of momentum conservation in the interaction.  The LHS of (\ref{eem}) describes, therefore, the change of the electron momentum at a point (the electron instantaneous position) while the RHS  describes the momentum carried away by the electron-self-field which is distributed over the whole space. This is a consequence of the imposed dichotomic treatment: while the electron is described as a discrete and well localized object, a particle, its self-field is a non-localized object distributed over the points of the entire space and whose contribution to the changes in the electron must be computed from all these points. It introduces a strong non-locality and excludes the possibility of a true  equation of motion which would give an essentially local description. A true (in the sense of local) equation of motion for a classical charged particle is then possible only in the context of discrete interactions mediated by exchanged (classical) photons \cite{CPMF}. The RHS of (\ref{fld}) would be replaced in this case by the momentum of the emitted photon, while the RHS remains local as it is always defined at a single point, the electron position. The space-time average of this (then local) equation would reproduce (\ref{fld}). By the way, $F^{\mu\nu}_{ext}V_{\nu}$ in the LHS of (\ref{fld}) is the spacetime average of the momentum exchanged between the electron and the external charges while $\hbox{\Large a}^{\mu}$ is the electron average acceleration. In the context of the LWS (1), the equation (\ref{fld}) must, therefore, be regarded as an effective equation that would be better represented as
\begin{equation}
\label{finally}
m\hbox{\Large a}^{\mu}-F^{\mu\nu}_{ext}V_{\nu}=-<\frac{2}{3}\hbox{\Large
a}^{2}V^{\mu}>,
\end{equation}
where the bracketed term represents the contribution from the electron
self-field:
\begin{equation}
\label{70a}
 <\frac{2}{3}\hbox{\Large
a}^{2}V^{\mu}>=\lim_{{\varepsilon_{1}\to0}\atop{\varepsilon_{2}\to\infty}}\int
dx^{3}\nabla_{\nu}\Theta^{\mu\nu}\theta(\rho-\varepsilon_{1})\theta(\varepsilon_{2}-\rho).
\end{equation}
 This is more than just a change of notation; it explicitly implies on a clear distinction between the V inside and the V outside the bracket in (\ref{finally}): $$<V>\neq V.$$ This distinction between the LHS and the RHS of (\ref{eem}) is missing in equation (\ref{fld}); it was deleted by the integration process. It represents the strong non-locality introduced at the beginning with the hypothesis of a continuous interacting field (1).\\ It makes no sense, therefore, multiplying (\ref{fld}) or (\ref{finally}) by V. This would be a mixing of instantaneous and average values.  Using (\ref{fld}) or (\ref{finally}) for this does not make any sense One should instead try to follow the associated physical picture.
The LHS of (\ref{eem}) multiplied by V is null because the force that drives the electron with the 4-velocity V delivers a power  $m\hbox{\Large a}_{0}V^{0}$) that is equal to
the work per unit time ($m\vec{\hbox{\Large a}}.\vec{V}$) realized by this force along the $\vec{V}$ direction
 (this, its well known, is the physical meaning of
$m\hbox{\Large a}.V=0$).  But this reasoning does not apply to the RHS of
(\ref{eem}) multiplied by V because the flux of  radiated energy is through a
spherical  surface $\rho=\varepsilon_{2},$ along K at each point,  not along $V$ (except at $\rho=0$, because of (\ref{limitK})); in order to make sense, as one is doing a balance of the flux-rate of energy, one has to add this flux-rate from each point  of the integration domain. Based on considerations of symmetry one can anticipate that the final result must be null: to each point of a spherical hypersurface $\rho=const.,\;\tau=\tau_{2},$ that gives a non-null contribution there is another point giving an equal but with opposite sign contribution. The RHS of (\ref{fld}) cannot be used for this point-to-point calculation as it just represents a kind of average or resulting value. For doing this balance one must start again from the beginning.  Contributions from the electron self-field must always be calculated through this point-by-point summation, like in the  RHS of (\ref{eem}) for the flux of electromagnetic energy-momentum, through the walls of a Bhabha tube around the charge world-line, in the limit of $\rho\rightarrow0$. In particular, 
\begin{equation}
\label{dKt}
\int_{\tau_{1}}^{\tau_{2}} d\tau({m\hbox{\Large
a}}-V.F_{ext}).V=-\lim_{{\varepsilon_{1}\to0}\atop{\varepsilon_{2}\to\infty}}\int d^{4}x  X_{\mu}\nabla_{\nu}\Theta^{\mu\nu}\theta(\varepsilon_{2},\rho,\varepsilon_{1})\theta(\\tau_{2},\tau,\tau_{1}),
\end{equation}
where
\begin{equation}
\label{casesX}
X=\cases{K,&if $\rho>0$;\cr
V,&if $\rho=0$.\cr}
\end{equation}
X, in the RHS of (\ref{dKt}), gives the direction of the flux-rate of the
radiated energy; in the LHS this direction is given by
V.
Observe that $X(\tau_{ret})$ is x-dependent and so it does not commute with
$\int d^{4}x$, that is, X inside and X outside the integral in the RHS of
(\ref{dKt}) give distinct results and, based on the above arguments, one is
saying that (\ref{dKt}) shows the correct way.
Its LHS is, of course, null. It will be shown now that the RHS is also null, so that
there is no contradiction anymore.  One knows that
\begin{equation}
\label{Vdt1}
\nabla_{\nu}\Theta^{\mu\nu}=\frac{1}{4\pi}F^{\mu}_{\;\alpha}\;\nabla_{\nu}F^{\alpha\nu}=\frac{1}{4\pi}F^{\mu}_{\;\alpha}\;\Box A^{\alpha},
\end{equation}
and by direct calculation  one finds that
\begin{equation}
\label{boxa}
\Box A^{\mu}=\frac{K^{2}}{\rho^{3}}{\Big(}3\rho E\hbox{\Large
a}^{\mu}+\rho^{2}{\dot{\hbox{\Large
a}}}^{\mu}+(3E^{2}+\rho^{2}{\dot{\hbox{\Large a}}}_{K})V^{\mu}{\Big )}.
\end{equation}
So, the integrand of the RHS of (\ref{dKt}) is null for $\rho>0$
as $K^{2}=0$ there. For simplicity one could then just have used V instead of X in (\ref{dKt}), but see the next section for an alternative illuminating
calculation. Therefore, one just has to verify that
$\rho^{2}V_{\mu}\nabla_{\nu}\Theta^{\mu\nu}{\big|}_{\rho=0}$ is finite, or
equivalently that 
$\rho^{3}V_{\mu}\nabla_{\nu}\Theta^{\mu\nu}{\big|}_{\rho=0}=0.$
As 
\begin{equation}
V_{\mu}F^{\mu\nu}=\frac{1}{\rho^{2}}(EK^{\alpha}-W^{\alpha}),
\end{equation}
then
\begin{equation}
\label{Vdt}
4\pi\rho^{5}V_{\mu}\nabla_{\nu}\Theta^{\mu\nu}=-K^{2}{\Big(}2E\rho^{2}\hbox{\Large a}^{2}+3E(1-E^{2})+\rho^{2}(\rho{\dot{\hbox{\Large a}}}.\hbox{\Large a}-E{\dot{\hbox{\Large a}}}_{K}){\Big )},
\end{equation}
and
\begin{equation}
\label{last}
\lim_{\rho\to0}\rho^{3}V_{\mu}\nabla_{\nu}\Theta^{\mu\nu}=\lim_{\rho\to0}\frac{R^{2}{\Big(}2E\rho^{2}\hbox{\Large a}^{2}+3E(1-E^{2})+\rho^{2}(\rho{\dot{\hbox{\Large a}}}.\hbox{\Large a}-E{\dot{\hbox{\Large a}}}_{K}){\Big)}}{\rho^{4}}.
\end{equation}
And this is null at the limit $\rho\rightarrow0$ because $${\cal O}[2E\rho^{2}\hbox{\Large a}^{2}+3E(1-E^{2})+\rho^{3}{\dot{\hbox{\Large
a}}}.\hbox{\Large a}-\rho E{\dot{\hbox{\Large a}}}_{R}]+{\cal O}[R^{2}]=3+2>4,$$ according to (\ref{NP}). So, both sides of (\ref{dKt}) are
equally null and there is no contradiction. This is in agreement with the fact
that due to (\ref{BoxA},\ref{J}) and to the antisymmetry of F,
$$V_{\mu}\nabla_{\nu}\Theta^{\mu\nu}=\frac{1}{4\pi}V_{\mu}F^{\mu}_{\;\;\;\alpha}\nabla_{\nu}F^{\alpha\nu}=V_{\mu}F^{\mu}_{\;\;\alpha}J^{\alpha}=0.$$
\begin{section}
{ Using the divergence theorem}
\end{section}

For the sake of a better understanding of the meaning of X in eq. (\ref{dKt}) 
its RHS will be worked out with the use of the divergence theorem. Then one has
\begin{eqnarray}
\lefteqn{\int_{\tau_{1}}^{\tau_{2}} d\tau({m\hbox{\Large
a}}-V.F_{ext}).V=
\lim_{{\varepsilon_{1}\to0}\atop{\varepsilon_{2}\to\infty}}\int
dx^{4}{\Big\lbrace}\Theta^{\mu\nu}\nabla_{\nu}X_{\mu}\;\theta(\varepsilon_{2},\rho,\varepsilon_{1})\theta(\tau_{2},\tau,\tau_{1})+}
\hspace{15ex}\nonumber\\
& &                     
+X_{\mu}\Theta^{\mu\nu}{\Big[}\nabla_{\nu}\rho{\Big(}\delta(\rho-\varepsilon_{1})\theta(\varepsilon_{2}-\rho)-\theta(\rho-\varepsilon_{1})\delta(\varepsilon_{2}-\rho){\Big)}\theta(\tau_{2},\tau,\tau_{1})\nonumber\\ 
& & -K_{\nu}{\Big(}\delta(\tau-\tau_{1})\theta(\tau_{2}-\tau)-\theta(\tau-\tau_{1})\delta(\tau_{2}-\tau){\Big)}\theta(\varepsilon_{2},\rho,\varepsilon_{1}){\Big]}{\Big\rbrace},\label{X}
\end{eqnarray}

The explicit dependence on $\nabla_{\nu}X_{\mu}$ makes clear why one cannot just
use K instead of X in (\ref{dKt}): although $\lim_{\rho\to0} K = V$,
$\lim_{\rho\to0} \nabla K \ne \nabla V= -K\hbox{\Large a}.$\\
For working out the first term of the RHS of (\ref{X}) one needs (\ref{nablarho2}-\ref{Knro}) and
\begin{equation}
\nabla_{\mu}K_{\nu}=\nabla_{\mu}(\frac{R_{\nu}}{\rho})=\frac{\eta_{\mu\nu}+K_{\mu}V_{\nu}}{\rho}-\frac{K_{\nu}}{\rho}\nabla_{\mu}\rho,
\end{equation}
\begin{equation}
\Theta^{\mu\nu}\eta_{\mu\nu}=0.
\end{equation}
Then, from (\ref{t'}) and $K^{2}=0$ one has for the upper limit
\begin{equation}
\label{upper}
\lim_{\varepsilon_{2}\to\infty}\int^{\varepsilon_{2}}
dx^{4}\Theta^{\mu\nu}\nabla_{\nu}K_{\mu}=\lim_{\varepsilon_{2}\to\infty}\int d\tau\int^{\varepsilon_{2}}\frac{d\rho}{\rho^{3}}=0.
\end{equation}
For the lower limit $\nabla_{\nu}X_{\mu}=\nabla_{\nu}V_{\mu}=-K_{\nu}\hbox{\Large a}_{\mu}$ and
then, from (\ref{t'}),
\begin{equation}
4\pi\rho^{4}K.\Theta.\hbox{\Large a}=\hbox{\Large
a}_{K}(K^{2}W^{2}-1)=\hbox{\Large a}_{K}(K^{2}+1)+\rho^{2}\hbox{\Large
a}_{K}K^{2}(\hbox{\Large a}^{2}-\hbox{\Large a}_{K}^{2})-\rho\hbox{\Large
a}_{K}^{2} K^{2}
\end{equation}
So,
\begin{equation}
\label{lower}
\lim_{\rho\to0}4\pi\int\rho^{2}d\rho K.\Theta.\hbox{\Large
a}=-\frac{\hbox{\Large a}_{K}(K^{2}+1)}{\rho} +\hbox{\Large
a}_{K}K^{2}(\hbox{\Large a}^{2}-\hbox{\Large a}_{K}^{2})\rho-\hbox{\Large
a}_{K}^{2}K^{2}ln\rho=0,
\end{equation}
because
\begin{equation}
\lim_{\rho\to0}\frac{\hbox{\Large
a}_{K}(K^{2}+1)}{\rho}=\lim_{\rho\to0}\frac{\hbox{\Large
a}_{R}(R^{2}+\rho^{2})}{\rho^{4}}=0,
\end{equation}
as
\begin{equation}
{\cal O}[\hbox{\Large a}_{R}]+{\cal O}[R^{2}+\rho^{2}]=2+4>4,
\end{equation}
and
\begin{equation}
\lim_{\rho\to0}\hbox{\Large a}_{K}K^{2}(\hbox{\Large a}^{2}-\hbox{\Large
a}_{K}^{2})\rho=0.
\end{equation}
For evaluating the limit of the last term of (\ref{lower}) one has considered that 
$$K^{2}\hbox{\Large a}_{K}^{2}=\frac{R^{2}(\hbox{\Large a}.R)^{2}}{\rho^{4}}$$
\begin{equation}
{\cal O}[K^{2}\hbox{\Large a}_{K}^{2}]=2={\cal O}[\rho^{2}]
\end{equation}
 to see that
\begin{equation}
\lim_{\rho\to0}\hbox{\Large a}_{K}^{2}K^{2}ln\rho\sim
\lim_{\rho\to0}\rho^{2}ln\rho=0.
\end{equation}
It is important to use the appropriate values of X to have consistent results.
The use, for example, of $X=V$  in the upper limit or of $X=K$ in the lower
limit would produce inconsistent results.
The second line of (\ref{X}) is composed of two terms, with $\rho=\varepsilon_{1}$ and $\rho=\varepsilon_{2},$ respectively.
For the $\rho=\varepsilon_{1}$ term, $X=V$ in the limit and then one has from (\ref{t'}) that
\begin{equation}
4\pi\rho^{4}V.\Theta.\nabla\rho=(1+K^{2}E)(W^{2}+E)+\frac{\rho\hbox{\Large
a}_{K}}{2}(1-K^{2}W^{2}).
\end{equation}
Therefore,
\[\lim_{\rho\to0}4\pi\rho^{2}V.\Theta.\nabla\rho   =
\lim_{\rho\to0}{\Big(}\frac{(\rho^{2}+R^{2}+R^{2}\hbox{\Large
a}_{R})(\rho^{2}\hbox{\Large a}^{2}-\hbox{\Large a}_{R}^{2}-\hbox{\Large
a}_{R})}{\rho^{4}}+ \]
\begin{equation}
 \mbox{}+ \frac{{\hbox{\Large a}_{R}}[\rho^{2}+R^{2}-R^{2}(\rho^{2}\hbox{\Large
a}^{2}-\hbox{\Large a}_{R}^{2}-2\hbox{\Large a}_{R})]}{2\rho^{4}}{\Big)}=0,
\end{equation}
because
\begin{equation}
{\cal O}[\rho^{2}+R^{2}+R^{2}\hbox{\Large a}_{R}^{2}]+{\cal
O}[\rho^{2}\hbox{\Large a}^{2}-\hbox{\Large a}_{R}^{2}-\hbox{\Large
a}_{R}]=4+2>4
\end{equation}
and
\begin{equation}
{\cal O}[\hbox{\Large a}_{R}]+{\cal O}[(\rho^{2}+R^{2})-R^{2}(\rho^{2}\hbox{\Large a}^{2}-\hbox{\Large
a}_{R}^{2}-2\hbox{\Large a}_{R})]=2+4>3
\end{equation}
Again one  has consistent results only if one uses the correct values of X in its
respective limiting situation.\\
For the $\rho=\varepsilon_{2}$ term one has $X=K$ and then $$4\pi\rho^{2}K.\Theta.\nabla\rho=\frac{1}{2}\frac{K.\nabla\rho}{\rho^{2}}=-\frac{1}{2\rho^{2}}.$$ So, it is null in the limit when $\rho$ tends to $\infty.$\\
Finally, the third line of (\ref{X}) does not contribute because $\rho^{2}K.\Theta.K\equiv0$ for $\rho>0$ and produces a finite result at the limiting $\rho=0,$ or equivalently: $$lim_{\rho\rightarrow0}\rho^{3}K.\Theta.K=0.$$

\begin{section}
{ THE MEANING OF THE CLASSICAL FIELD}
\end{section}
An strict observance of the two geometric constraints ($R^{2}=0$ and $R.dR=0$) in the LWS allows the introduction of the extended causality, the interpretation of theses solutions in terms of creation and annihilation of particles, and the vision of the  interacting electromagnetic field as composed of discrete point-like objects, the classical photons. The continuous picture of a wave and the idea of its continuous emission are just approximations valid for large distance and macroscopic sources; it is justified for the normally large number of photons involved. The short distance limit of CED is drastically changed with the extended causality: old inconsistencies, like the non-integrability of the self-field energy tensor, disappears. The paradoxes associated to the electron equation of motion are all explained with the understanding that this is an effective equation, written in terms of averaged values and, therefore, limited on its applications and validity.
The implicit non-locality of (\ref{finally}) is a consequence of the explicit bi-locality of (\ref{LWS}) and consequently of its energy tensor $\Theta$: they both depend on $R=x-z(\tau_{ret}).$ Although $\varepsilon_{1}\rightarrow0$ in (\ref{finally}), the term $<\frac{2}{3}\hbox{\Large
a}^{2}V^{\mu}>$ comes from the limiting $\varepsilon_{2}\rightarrow\infty.$
 The electromagnetic fundamental (in the sense of irreducible) interaction is the exchange of a single photon.  In classical physics this can be seen as the intersection of A(x) with the light-cone generator that connects the two charges, as depicted in figure 3, and not by A(x) itself, which rather represents the smearing of this interaction on the charge lightcone. That is why a space integration is necessary to retrieve the momentum carried out by the photon, which is the meaning of the RHS of equations (\ref{eem}) and (\ref{ef}). Their RHS's are a point-by-point summation of the contribution of each light-cone generator. Equation (\ref{fld}) cannot therefore be a true equation of motion because it does not describe the single photon exchange, as required by the fundamental interaction, but (an average) of all photons inside the integration domain. This explains why (\ref{fld}) is not time-reversal invariant as a fundamental equation must be.
The energy flux from the charge is, of course, non null at the point $z(\tau_{ret})$, as it is detected at x, but it is indeed null at $z(\tau_{ret}\pm d\tau)$ as one concludes from (\ref{fpe}) and (\ref{5eflc}).  This has some noticeable consequences. It shows that the classical radiation process is discrete in time; this discreteness takes the $\rho\rightarrow0$-limit to be revealed. At $\rho>0$ this effect is masqueraded by the average character of A(x). It is also in direct contradiction to the Gauss's law, which makes no sense if the field is seen as the effect of a discrete exchange of particle-like objects, unless the field is taken as the average of these effects in space and time. It requires, therefore, a re-avaluation of 
the physical meaning of this law and of the Faraday-Maxwell concept of fields. 
This question is also relevant to quantum theories (Quantum Mechanics and Quantum Field Theory) because it deals with the faithfulness of the interaction description. How far does the classical field (that, in QFT, one wants to quantize) really represents the experimentally observed interactions? 
This is closely related to the distinct contents of the Coulomb's law and of the Gauss' law. While the first one gives a strict description of what is actually observed, i.e., a force between two charges, acting on each one along the straightline connecting them, the second one contains an extra assumption (the Faraday-Maxwell concept of field) that effectively {\bf extends} this effect, observed at the charge position only, to all points in the space surrounding the charge, regardless the presence or not of the second charge.\\ The concept of a field existing everywhere around a single charge, regardless the presence of any other charge is an extrapolation of what is effectively observed. There is, therefore, a very deep distinction between the Coulomb's and the Gauss's laws.
This last one describes the {\it inferred} electric field as existing around a single charge, independent of the presence of the other charge.
The electric field, as it is well known, is extracted from the Gauss's law
through the integration of its flux across a closed surface, having the appropriate symmetry, {\it enclosing} the charge,
\begin{equation}
\label{Gauss}
{\vec E}(x)={\hat n} \frac{\int^{x}_{V}\rho dv}{\int_{\partial V}dS},
\end{equation}
where ${\hat n}$ is the unit vector normal to the surface $\partial V.$
The eq. (\ref{Gauss}) puts in evidence the effective or average character of
the Maxwell's concept of field; it gives also a hint on the meaning and origin of the field singularity. If the electric field can be visualized in terms of exchanged photons, then according to (\ref{Gauss}), the frequency or the number of these exchanged photons must be proportional to the enclosed net charge. And if we take ${\vec E}$, as suggested by the Gauss' law, as a
measure of the average number of photons emitted/absorbed by a point charge, we can schematically write, $E\sim \frac{n}{4\pi r^{2}},$ where n is the number of photon per unit of time crossing an spherical surface of radius r and centred on the charge. Then, the divergence of E  when $r\rightarrow0$ does not represent a physical fact like an increasing number of photons, but just an increasing average number of photons per unit area, as the number of photons remains constant but the area tends to zero. So, a field singularity would have no physical meaning, it would just be a consequence of this average nature of the Maxwell's field.\\ In the modern perspective of seing a fundamental electromagnetic interaction as the result of a single photon exchange, the classical electromagnetic field describes rather  the smearing of this interaction in the the time and in the space around each charge. No wonder one finds inconsistencies in the theory short distance limit if one is replacing the interaction by its space average.

%
\begin{center}
{LIST OF FIGURE CAPTIONS}
\end{center}
\begin{enumerate}
\item Fig. 1.{The usual interpretation of the Lienard-Wiechert solutions. By the point x passe two spherical waves: the retarded one, created in the past $\tau_{ret}$, and the advanced one, created in the future $\tau_{adv}.$ J is the source of both.}
\item Fig. 2.{Creation an annihilation of particle in classical physics as a new interpretation of the LWS. At x there are two (classical) photons. One, created in the past by J, at $\tau_{ret},$ and propagating along the lightcone generator K.  J is its source. The other one, propagating along ${\bar K}$, will be absorbed in the future by J, at $\tau{adv}.$ J is its sink. Both are retarded and pointlike solutions.}
\item Fig. 3.{Double limiting process:$\rho\rightarrow0$ along K and
$\tau\rightarrow\tau_{ret}$. The single limiting process $x\rightarrow z(\tau_{ret})$ along K (or $\rho\rightarrow0$) does not solve the $(\frac{0}{0})$-indeterminacy in the definition of K at $\rho=0.$ $K=\frac{x-z(\tau_{ret})}{\rho}$ for $\rho>0.$ It takes a second simultaneous limit $\tau\rightarrow\tau_{ret}$ along the electron worldline.}
\item Fig. 4.{Classical picture of the fundamental quantum process: at $\tau_{ret}$ an  electron with a 4-velocity $V_{1}$, changes it to $V_{2}$ by  emitting a photon with a 4-velocity K.  $\tau_{ret}$ is a singular point  on the electron world-line because of this indeterminacy on its tangent. No infinity is involved.}
\end{enumerate}
%
\begin{figure}
\epsfxsize=200pt
\epsfbox{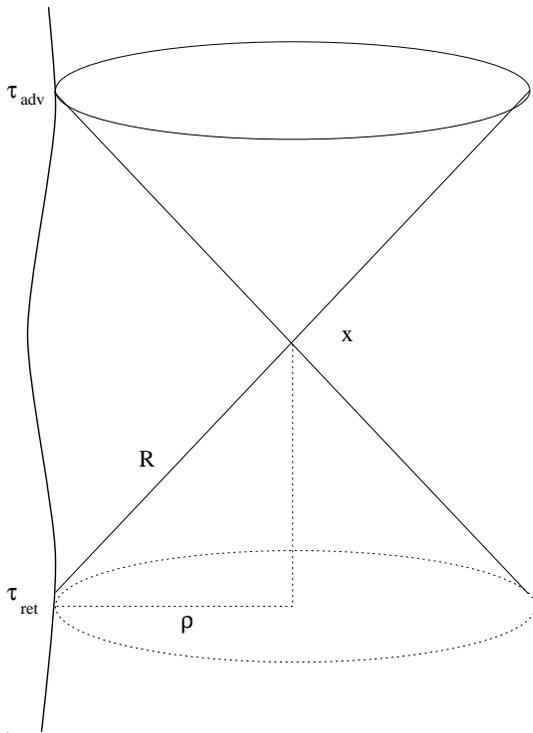}
\caption[Fig. 1.]{The usual interpretation of the Lienard-Wiechert solutions. By the point x passe two spherical waves: the retarded one, created in the past $\tau_{ret}$, and the advanced one, created in the future $\tau_{adv}.$ J is the source of both.}
\end{figure}
%
%
\begin{figure}
\epsfbox{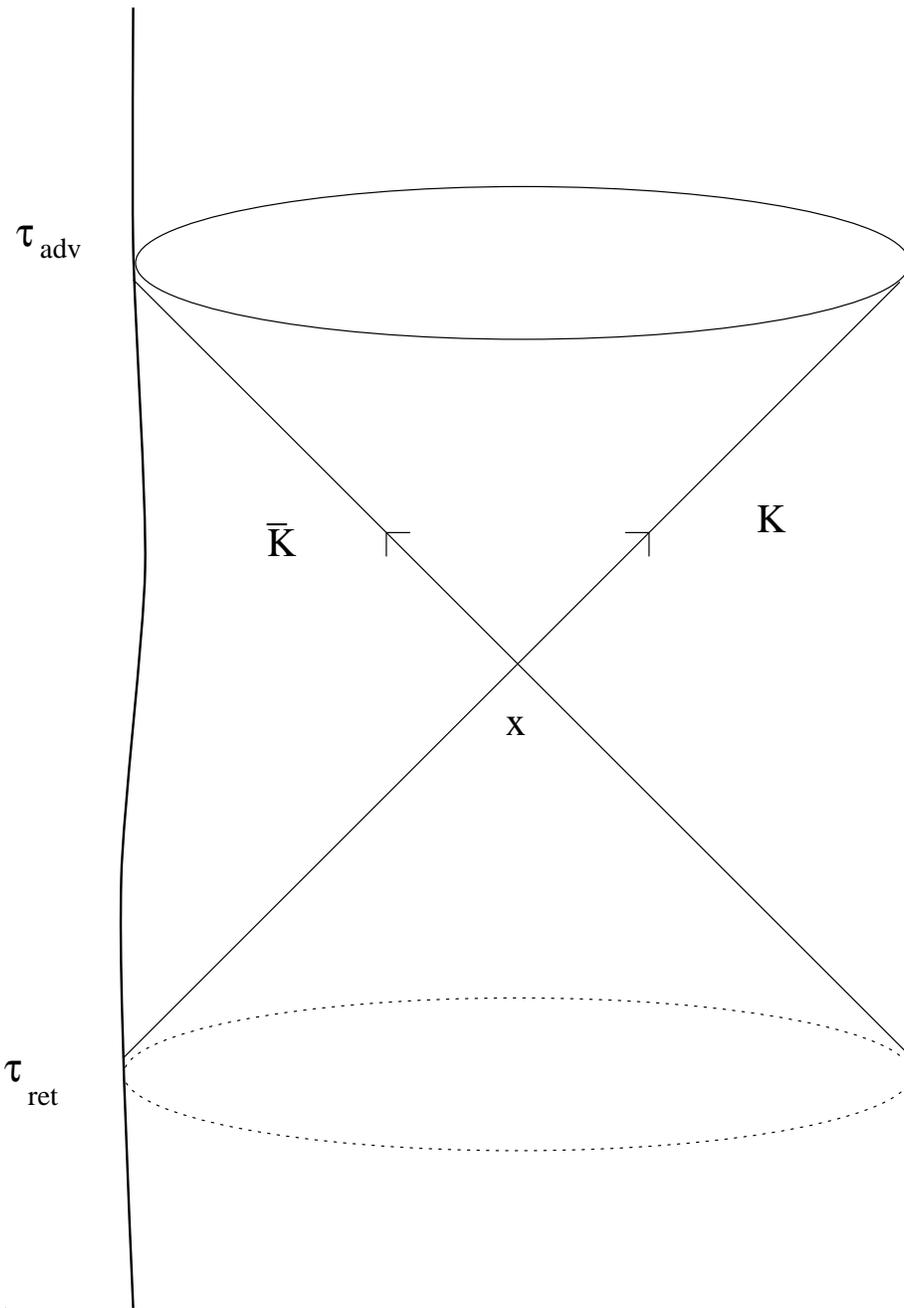}
\epsfxsize=100pt
\caption[Fig. 2.]{Creation an annihilation of particle in classical physics as a new interpretation of the LWS. At x there are two (classical) photons. One, created in the past by J, at $\tau_{ret},$ and propagating along the lightcone generator K.  J is its source. The other one, propagating along ${\bar{K}}$, will be absorbed in the future by J, at $\tau_{adv}.$ J is its sink. Both are retarded and pointlike solutions.}
\end{figure}
%
%
\begin{figure}
\epsfxsize=200pt
\epsfbox{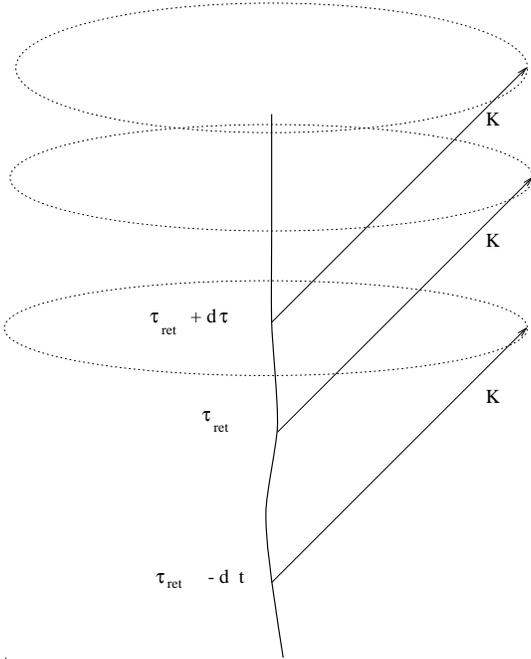}
\caption[Fig. 3.]{Double limiting process:$\rho\rightarrow0$ along K and
$\tau\rightarrow\tau_{ret}$. The single limiting process $x\rightarrow z(\tau_{ret})$ along K (or $\rho\rightarrow0$) does not solve the $(\frac{0}{0})$-indeterminacy in the definition of K at $\rho=0.$ $K=\frac{x-z(\tau_{ret})}{\rho}$ for $\rho>0.$ It takes a second simultaneous limit $\tau\rightarrow\tau_{ret}$ along the electron worldline.}
\end{figure}
%
%
\begin{figure}
\epsfxsize=200pt
\epsfbox{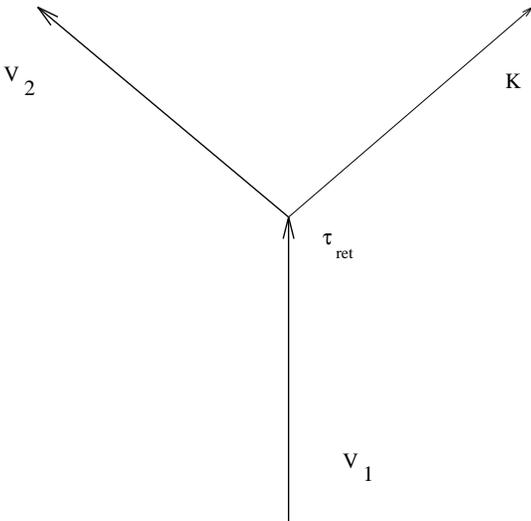}
\caption[Fig. 4.]{Classical picture of the fundamental quantum process: at $\tau_{ret}$ an  electron with a 4-velocity $V_{1}$, changes it to $V_{2}$ by  emitting a photon with a 4-velocity K.  $\tau_{ret}$ is a singular point  on the electron world-line because of this indeterminacy on its tangent. No infinity is involved.}
\end{figure}
\end{document}